\documentclass{ab}

\usepackage{graphicx}
\usepackage{natbib,longtable}

\def\degr{\hbox{$^\circ$}}

\begin{document}

\title{Are polar rings indeed polar?}

\author{K.~I.~Smirnova$^1$ and A.~V.~Moiseev$^2$}

\institute{$^1$Ural Federal University, Ekaterinburg, 620002 Russia\\
$^2$Special Astrophysical Observatory, \frame{Russian Academy of Sciences}\thanks{The system of Russian Academy of Sciences institutes was liquidated (``re-organized'') on Sep 2013},
Nizhnij Arkhyz, 369167, Russia}

\titlerunning{Are polar rings indeed polar?}
\authorrunning{Smirnova \& Moiseev}%

\date{February 13, 2012/Revised: February 20, 2012}
\offprints{K. Smirnova \email{arashu@rambler.ru}, A. Moiseev \email{moisav@sao.ru}}

\abstract{ We have considered polar ring galaxy candidates, the images of
which can be found in the SDSS. The sample of 78 galaxies
includes the most reliable candidates from the SPRC and PRC
catalogs, some of which already have kinematic confirmations. We
analyze the distributions of studied objects by the angle between
the polar ring and the central disk, and by the optical diameter
of the outer ring structures. In the vast majority of cases, the
outer structures lie in the plane close to polar \mbox{(within
$10\degr$--$20\degr$)} which indicates the stability of the
corresponding orbits in the gravitational potential of the halo.
Moderately inclined outer structures are observed only in about
$6\%$ of objects which probably indicates their short lifetime. In
such an unstable configuration, the polar ring would often cross
the disk of the galaxy, being smaller than it in the diameter. We
show that the inner polar structures and outer large-scale polar
rings form a single family in the distribution of diameters
normalized to the optical size of the galaxy. At the same time,
this distribution is bimodal, as the number of objects with
\mbox{$d_{\rm ring}= (0.4$--$0.7)\,d_{\rm disk}$} is negligible. Such a
shape of size distribution is most likely due to the fact that the
stability of polar orbits in the inner regions of galaxies is
maintained by the bulge or the bar, while in the outer regions it
is provided by the spheroidal (or triaxial) halo.
}


\maketitle

\section{INTRODUCTION}
\label{intro}

Polar ring galaxies (PRG) simultaneously demonstrate rotation
about two axes: apart from the central stellar disk,  an extended
structure composed of stars and gas, called a polar ring or a
polar disk, rotates in the plane perpendicular or highly inclined
to it. Prior to the release of our paper \citep{Moiseev_SDSS}, virtually all the studies of
such galaxies referred to the objects presented in the catalog of
\citet{Whitmore1990}. This catalog
(Polar Rings Catalog, PRC), based on the study of photographic
data, is divided into four parts and includes 157 objects.
However, only six of them \mbox{(``type A'')} were kinematically
confirmed at the moment of publication. The authors have
attributed 27~galaxies to the type B (good candidates). A major
part of the catalog was comprised of peculiar objects with various
signs of interaction with the environment, only possibly related
to the phenomenon of polar rings.

To relegate a galaxy to the class of PRGs, the following has to be
fulfilled according to \citet{Whitmore1990}:
\begin{itemize}
\item the presence  in the galaxy of two sub-systems rotating in
almost orthogonal planes and having comparable rotation
velocities; \item both kinematically isolated subsystems should
have matching systemic velocities, i.e., to be an object with a
common center rather than a random projection of the two galaxies
on the line of sight, their photometric centers evidently have to
match; \item the polar structure (a ring or a disk) has to be
comparable in size with the central galaxy and look relatively
flat.
\end{itemize}

During the subsequent years of study of the  PRC objects, only
about two dozens of cases obtained kinematic confirmations that
these galaxies indeed belong to the PRGs. This means that the
presence of matter in polar orbits was verified using optical
spectroscopy and radio interferometry in the line of~H\,I.

\begin{figure}
\centerline{
\includegraphics[width=4cm]{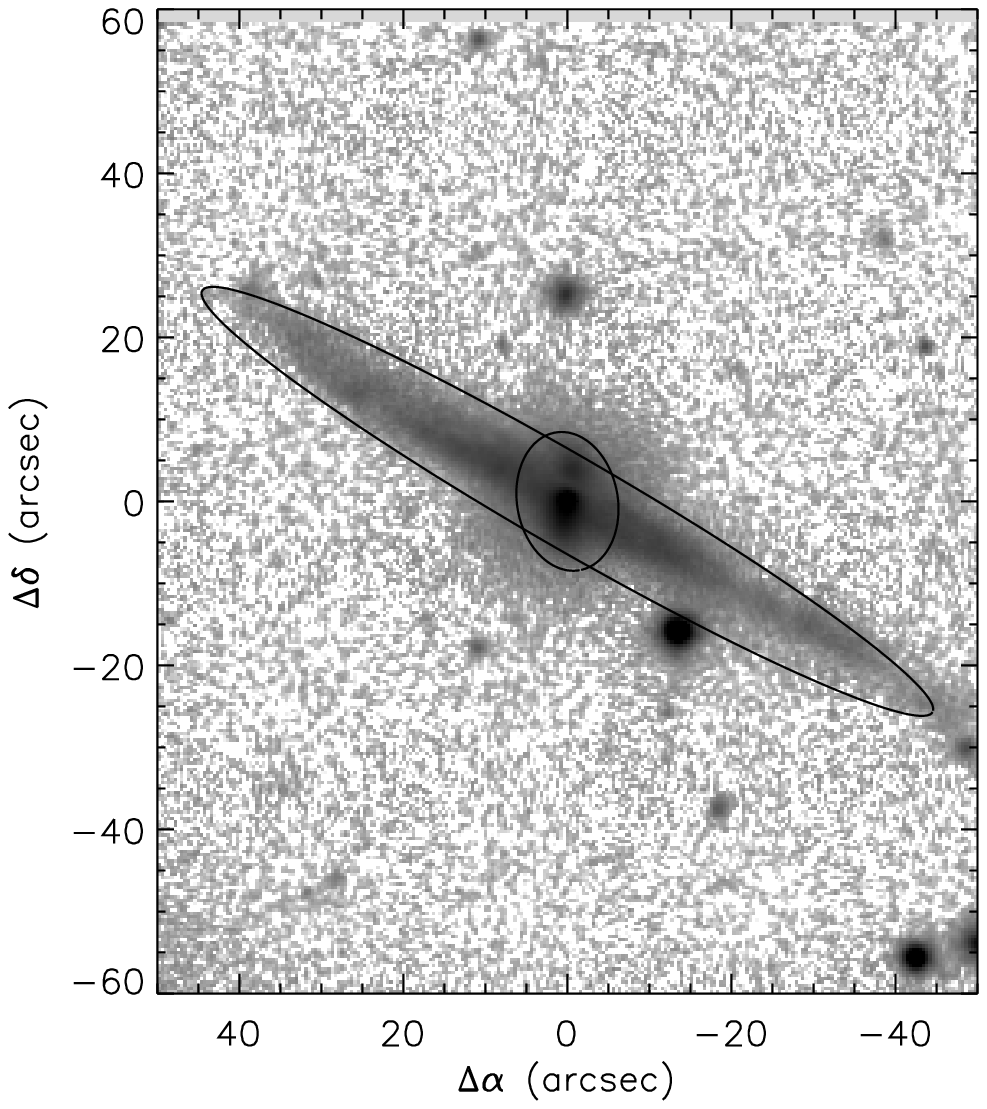}
\includegraphics[width=4cm]{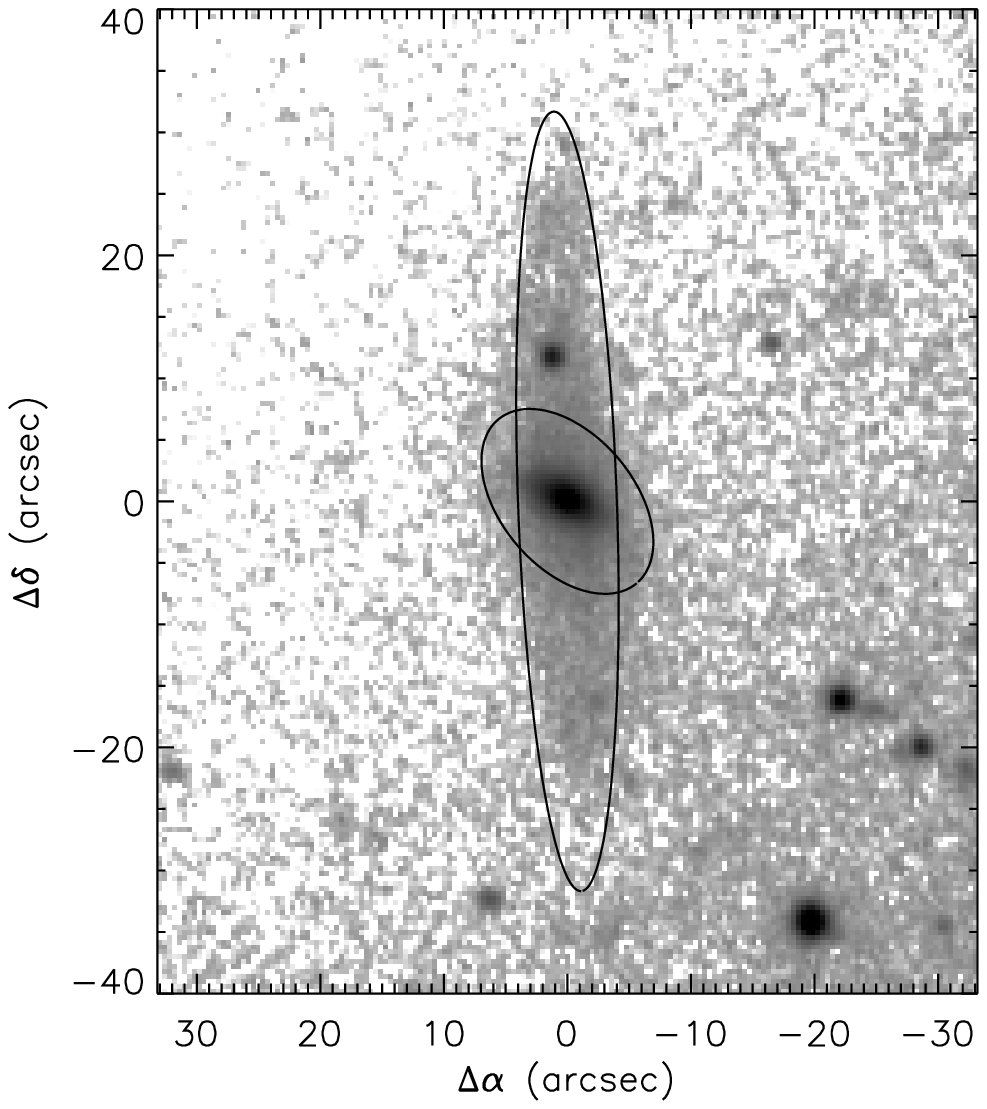}
}\caption{Image of the SPRC\,42 (left) and
SPRC\,50 (right) galaxies, the sum of five SDSS filters in the
logarithmic intensity scale. The ellipses show the accepted
orientations of the outer regions of the disk and the polar
structure.} \label{fig_1}
\end{figure}

In 2011 we have published  a catalog
of the polar ring candidates \citep{Moiseev_SDSS}, consisting of~275 candidates
selected from the images of the Sloan Digital Sky Survey (SDSS
DR7) mainly owing to the use of GalaxyZoo project
data\footnote{\tt www.galaxyzoo.org}. In this catalog
(\mbox{SDSS-based} Polar Ring Catalogue, SPRC), we also applied a
conventional separation of objects into four types based on their
optical morphology only:
\begin{itemize}

\item  the best candidates  (70 galaxies);

\item good candidates, including possible random projections (115
galaxies);

\item  related objects with strongly warped disks \mbox{(53
galaxies)};

\item rings viewed nearly face-on (37~galaxies).
\end{itemize}

Several scenarios of the formation of PRGs were developed. All of
them are in one or another way related with the interaction of the
central galaxy with its environment
(see~\citet{Combes03,Combes06} and references
therein). One scenario is a merger of two galaxies with mutually
orthogonal disks that occurs at relatively low mutual velocities.
Calculations show that a large part (about~3/4) of polar rings is
most likely produced from the accretion by a large galaxy of the
matter of its gas-rich companion having the corresponding
direction of its orbital rotation momentum. This requires very
specific initial
conditions~\citep{ReshetnikovSotnikova1997,Combes03}.
The capture of gas itself may  naturally occur at any orbital
orientation of the ``donor'' companion. However, if the gas ring
formed this way is inclined at a small angle to the stellar disk
of the galaxy, then in a few revolutions under the influence of
the gravity of the disk, the ring will fall in the plane of the
galaxy due to precession. At the same time, the polar (i.e. almost
perpendicular) orientation of the disk and the ring is stable,
hence the ring will survive a lot of revolutions. An important
stabilizing role is played here by the dark halo when
it has a prolate or oblate shape.

The scenario of formation of some polar rings in the process of
gas accretion from the filaments of the intergalactic medium (the  ``cold accretion'') is recently gaining
popularity~\citep{Maccio2006,Brook2008,Spavone2010}.
This may explain the existence of the largest polar structures in
which the mass of the polar component is equal to or greater than
the mass of the central galaxy. It is important to note that  in
itself, cold accretion is a natural stage undergone by galaxies
when they are gaining their baryon mass. But only in some specific
cases it is possible to notice the traces of this process.

To date, PRGs are among the few types of objects allowing to
directly estimate the shape of the dark halo owing to the ability
to measure the rotation curve in two perpendicular planes.

In order to understand exactly what mechanism led to the formation
of a PRG in each particular case, as well as to address the issue
of stability of polar structures, it is useful to consider the
statistical distributions of their main parameters, such as the
relative size and inclination to the plane of the central disk.
Are only the orbits orthogonal to the disk long-lived? How often
do polar rings cross the stellar disk of the host galaxy? The
statistics of the parameters of polar components in galaxies was
earlier considered in~\citet{Whitmore1991} on a small
sample of 27 PRC galaxies. Now we can significantly increase the
number of studied objects, using a homogeneous enough set of
\mbox{observations---the} SDSS images. Our paper is dedicated to
this goal.

\begin{figure*}
\centerline{\includegraphics[width=8cm]{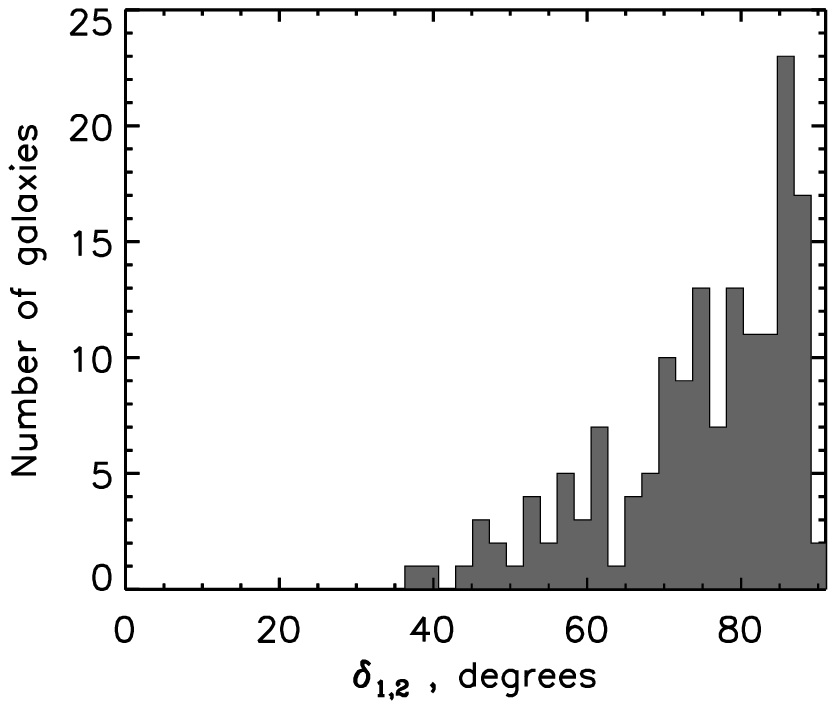}
\includegraphics[width=8cm]{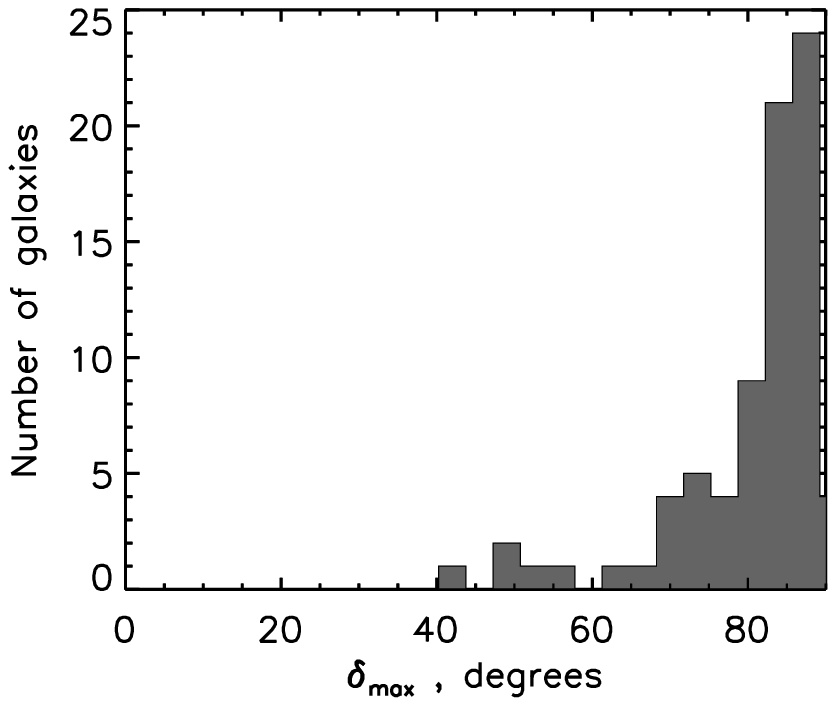}}
\caption{Histogram of distribution of the
angles of inclination of the polar structure to the galactic plane.
Left: all possible solutions by the formula
(\ref{eq2}). Right: only the maximum values of the
angles $\delta_{1,2}$ are taken.} \label{fig_2}
\end{figure*}

\begin{figure*}
\centerline{\includegraphics[width=8cm]{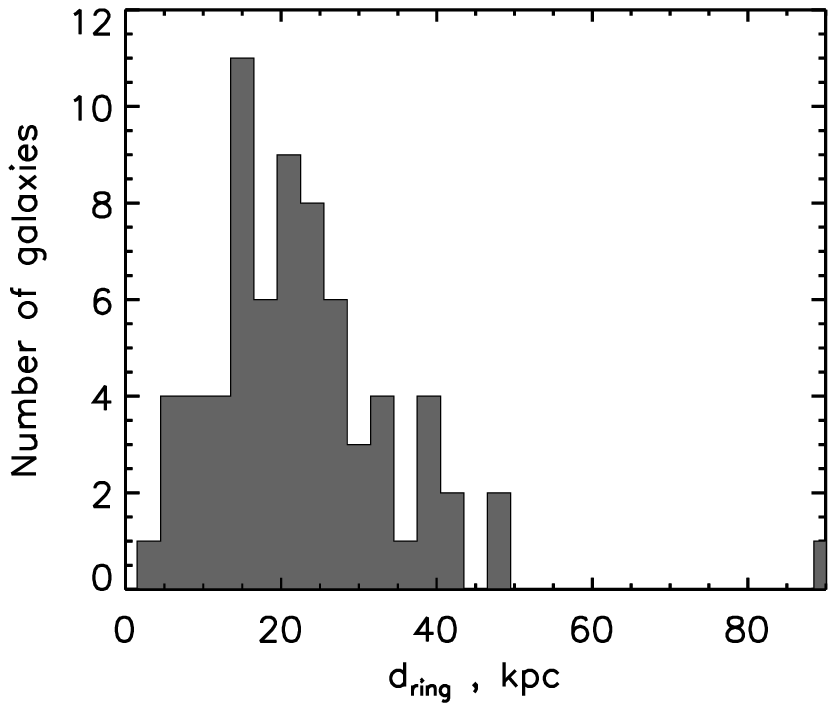}
\includegraphics[width=8cm]{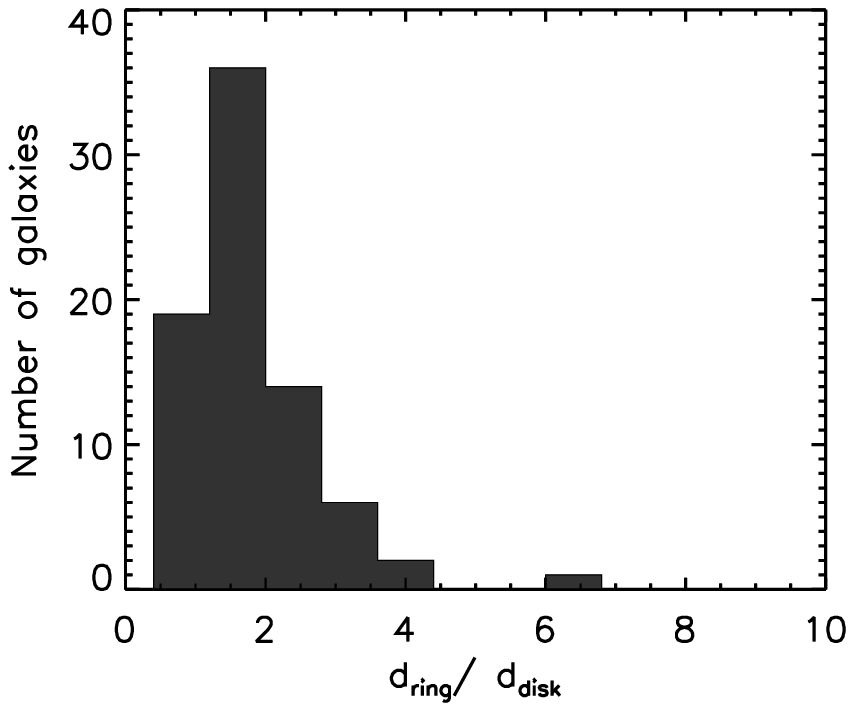}}
\caption{Distributions of the outer
diameters of polar structures. Left: linear size in kpc. A giant
ring SPRC\,50  with the diameter of 90~kpc stands out  from the
total distribution. Right: the diameter of the rings is normalized
to the size of the disk of the central galaxy.}
\label{fig_3}
\end{figure*}

\section{THE SAMPLE}

To study the statistics of the PRG parameters, it is important to
have the maximally uniform initial material. Our sample included
70 SPRC~galaxies classified by~\citet{Moiseev_SDSS} as
``the best candidates'' i.e., bearing the largest similarity
with classic PRGs. To these, we have added two galaxies classified
as ``rings oriented \mbox{face-on}'': SPRC\,241, SPRC\,260, which
have already been kinematically confirmed to belong to the
PRGs~\citep{Khoperskov12}. We have also included in the
sample the PRC catalog objects for which, on the one hand, there
are kinematic confirmations of belonging to the PRG  and, on
the other hand, there are SDSS/DR8 images. Such galaxies were only
six. In total, the sample consists of 78~galaxies with redshifts
in the range of    \mbox{$z= 0.003$--$0.17$}. It is important that
the analyzed images have similar limits of surface brightness,
since we use a common image source, SDSS/DR8 .

\section{ESTIMATION OF STRUCTURAL PARAMETERS}

In order to detect the features of low surface brightness of a
ring and a galaxy, we have summarized the images in the five SDSS
filters ($u,r,i,z,g$). Since the PRG images have complex shapes,
the use of standard automatic algorithms to fit the isophote
shapes is problematic. Therefore, we manually fitted the external
isophotes of the central disk and the polar ring by the ellipses,
using the tools of the {\tt SAOImage~DS9} software (see examples
in Fig.~\ref{fig_1}). The center of ellipses was fixed
on the core of the galaxy. Measured parameters are the major ($a$)
and minor ($b$) semiaxes of the ellipse and the position angle of
the major axis $\rm PA$. We also tried to describe the inner
borders of polar rings but managed to obtain a consistent result
for less than a dozen of galaxies, and hence the analysis of these
measurements was not done.  For the object SPRC\,33 (NGC\,4262),
which is a lenticular galaxy surrounded by a wide gaseous ring
hosting only a small number of stars visible mainly in the UV
range, the parameters of the ring and the galaxy were taken
from~\citet{Khoperskov2013}.

\subsection{Inclination Angles}

To determine the angle of inclination of the studied structures to
the line of sight ($i$), we used the classical formula for
estimating the oblate spheroid inclination from the apparent
axes ratio $(b/a)$, projected on the sky plane:
\begin{equation}
\cos^{2} i = \frac{(b/a)^{2} - q^2_0}{1-q^2_0},
\label{eq1}
\end{equation}
where the intrinsic flattening of the spheroid for \linebreak  the
central galaxies was assumed to be \linebreak
 \mbox{$q_0 = 0.20$}~\citep{Hubble_1926}, while
 the polar structures were considered to be flat:  \mbox{$q_0 = 0$}.

\begin{table*}
\caption{Parameters of galaxies}
\label{tab1}
\begin{tabular}{l|r|r|r|r|r|r|r|r|r|r}\hline\hline
Name    & \multicolumn{3}{c|}{Inner disk}	& \multicolumn{3}{c|}{ Polar ring} &     & &    &   \\	
   	    &a, $''$&b, $''$ &PA, $\degr$&a, $''$&b, $''$ &PA, $\degr$& z & $\delta_1$, $\degr$& $\delta_2$, $\degr$ & $D_{ring}/D_{disk}$\\	
\hline
  SPRC\,1&   7.2 &    3.6 &    92 &    5.6 &    2.4 &    10 &  0.073 &     71 &     84 &   0.78 \\
  SPRC\,2&  10.7 &    7.5 &    17 &   19.8 &    5.9 &    85 &  0.035 &     62 &     86 &   1.85 \\
  SPRC\,3&   5.7 &    2.8 &   325 &   12.1 &    2.9 &    66 &  0.037 &     86 &     74 &   2.11 \\
  SPRC\,4&   9.1 &    3.7 &   122 &    9.5 &    4.0 &    27 &  0.043 &     85 &     77 &   1.04 \\
  SPRC\,5&   4.0 &    3.2 &   220 &    8.7 &    2.0 &   106 &  0.028 &     86 &     65 &   2.20 \\
  SPRC\,6&   4.8 &    3.2 &    29 &    6.1 &    2.8 &   149 &  0.019 &     87 &     50 &   1.28 \\
  SPRC\,7&   4.2 &    3.2 &   150 &   17.8 &    9.5 &    50 &  0.060 &     72 &     60 &   4.28 \\
  SPRC\,8&   5.2 &    2.8 &   130 &    7.6 &    2.4 &    27 &     -- &     88 &     69 &   1.46 \\
  SPRC\,9&   5.7 &    1.9 &    10 &    5.5 &    1.4 &    94 &  0.145 &     80 &     88 &   0.97 \\
 SPRC\,10&   5.7 &    1.8 &   320 &    6.9 &    1.0 &    40 &  0.042 &     78 &     82 &   1.21 \\
 SPRC\,11&   6.7 &    3.2 &   125 &   10.4 &    3.5 &    30 &  0.066 &     85 &     77 &   1.56 \\
 SPRC\,12&   4.0 &    2.3 &   135 &    6.3 &    2.0 &    37 &  0.063 &     86 &     73 &   1.60 \\
 SPRC\,13&   8.3 &    3.2 &    15 &   13.1 &    2.4 &    97 &  0.032 &     79 &     86 &   1.57 \\
 SPRC\,14&   8.7 &    5.2 &   130 &   17.8 &    5.9 &    35 &  0.032 &     83 &     75 &   2.05 \\
 SPRC\,15&   9.9 &    4.4 &   144 &   11.9 &    4.0 &    22 &  0.034 &     71 &     53 &   1.20 \\
 SPRC\,16&   5.5 &    2.5 &   119 &   10.7 &    2.4 &    48 &  0.060 &     67 &     78 &   1.93 \\
 SPRC\,17&  10.0 &    4.4 &   100 &    7.6 &    2.4 &    10 &  0.026 &     82 &     82 &   0.76 \\
 SPRC\,18&   4.2 &    2.8 &    45 &    5.5 &    2.6 &   150 &  0.082 &     82 &     61 &   1.33 \\
 SPRC\,19&   4.8 &    2.4 &   109 &    5.9 &    1.8 &    12 &  0.105 &     87 &     75 &   1.25 \\
 SPRC\,20&   5.1 &    4.0 &     0 &   17.0 &    1.6 &    97 &  0.074 &     89 &     81 &   3.34 \\
 SPRC\,21&   3.6 &    2.2 &     9 &    4.8 &    2.0 &    99 &  0.081 &     75 &     75 &   1.33 \\
 SPRC\,22&   3.4 &    2.6 &    70 &    7.3 &    2.0 &   335 &  0.160 &     81 &     75 &   2.18 \\
 SPRC\,23&   9.9 &    6.7 &    65 &   13.1 &    5.2 &     4 &  0.028 &     53 &     85 &   1.32 \\
 SPRC\,24&   8.3 &    7.1 &    80 &   17.8 &    5.7 &   340 &  0.047 &     79 &     68 &   2.14 \\
 SPRC\,25&   3.6 &    2.0 &    89 &    7.9 &    1.6 &     4 &  0.073 &     79 &     88 &   2.22 \\
 SPRC\,26&   3.8 &    2.4 &   177 &    6.7 &    1.2 &    99 &     -- &     74 &     86 &   1.79 \\
 SPRC\,27&   5.9 &    2.7 &   125 &   14.3 &    2.2 &    30 &  0.048 &     89 &     81 &   2.40 \\
 SPRC\,28&   6.0 &    3.2 &    60 &    5.4 &    2.0 &   320 &  0.077 &     87 &     70 &   0.90 \\
 SPRC\,29&   5.6 &    4.0 &     0 &    7.9 &    2.0 &   116 &  0.047 &     82 &     61 &   1.43 \\
 SPRC\,30&   3.8 &    2.2 &    30 &    3.8 &    1.6 &   130 &  0.075 &     84 &     68 &   1.00 \\
SPRC\,31&   7.2 &    5.2 &   105 &   14.0 &    2.4 &   358 &  0.050 &     85 &     71 &   1.94 \\
 SPRC\,32&   5.3 &    3.0 &     5 &    4.8 &    1.3 &   103 &  0.035 &     88 &     75 &   0.89 \\
 SPRC\,33&  55.0 &   47.0 &     0 &  120.0 &   50.0 &    25 &  0.005 &     37 &     85 &   2.18 \\
 SPRC\,34&   5.3 &    3.0 &   343 &    9.9 &    1.6 &    73 &  0.081 &     85 &     85 &   1.85 \\
 SPRC\,35&   5.3 &    1.7 &   115 &    9.1 &    1.3 &    15 &  0.068 &     82 &     78 &   1.70 \\
 SPRC\,36&   5.9 &    2.4 &    15 &    4.4 &    1.6 &   115 &     -- &     88 &     73 &   0.73 \\
 SPRC\,37&   4.4 &    2.4 &   134 &    6.3 &    2.0 &    39 &  0.068 &     84 &     76 &   1.45 \\
 SPRC\,38&  11.1 &    4.8 &   115 &   15.8 &    5.9 &    12 &  0.039 &     87 &     70 &   1.43 \\
 SPRC\,39&   7.1 &    4.0 &   115 &   15.8 &    3.6 &    50 &  0.029 &     62 &     76 &   2.22 \\
 SPRC\,40&  47.5 &   13.1 &   105 &   26.9 &    5.9 &    57 &  0.004 &     46 &     53 &   0.57 \\
 SPRC\,41&   9.3 &    3.8 &   342 &    9.1 &    2.6 &   100 &  0.061 &     71 &     58 &   0.98 \\
 SPRC\,42&   8.5 &    6.1 &    10 &   51.5 &    5.6 &    60 &  0.023 &     58 &     67 &   6.04 \\
 SPRC\,43&   4.8 &    3.2 &    10 &    7.5 &    1.8 &    67 &  0.171 &     56 &     75 &   1.58 \\
 SPRC\,44&   5.9 &    2.4 &    49 &    5.2 &    2.6 &   320 &  0.113 &     79 &     80 &   0.87 \\
 SPRC\,45&   4.8 &    4.0 &     0 &    8.8 &    2.4 &    85 &  0.072 &     74 &     79 &   1.83 \\
 SPRC\,46&   4.9 &    2.0 &    72 &    4.8 &    1.6 &     7 &  0.128 &     60 &     75 &   0.96 \\
 SPRC\,47&  31.7 &    5.6 &   110 &   27.7 &    7.1 &   360 &  0.031 &     70 &     70 &   0.88 \\
 SPRC\,48&   8.7 &    5.2 &    23 &   13.9 &    3.6 &   123 &  0.056 &     89 &     73 &   1.59 \\
 SPRC\,49&   9.9 &    5.9 &   310 &   15.8 &    6.9 &    43 &  0.068 &     77 &     73 &   1.60 \\
 SPRC\,50&   8.7 &    5.3 &    40 &   31.7 &    4.0 &     2 &  0.078 &     45 &     56 &   3.64 \\
 SPRC\,51&   4.2 &    2.0 &     7 &    6.7 &    2.8 &    90 &  0.075 &     73 &     85 &   1.62 \\
 SPRC\,52&  21.6 &    4.6 &    80 &   15.8 &    4.0 &   146 &  0.015 &     65 &     67 &   0.73 \\
 SPRC\,53&   5.1 &    2.2 &   312 &    4.0 &    1.8 &    48 &  0.083 &     85 &     74 &   0.78 \\
 SPRC\,54&   5.3 &    4.3 &    45 &   14.6 &    2.2 &   334 &  0.039 &     71 &     85 &   2.74 \\
 SPRC\,55&   4.8 &    3.6 &     0 &   13.9 &    1.5 &    85 &  0.086 &     82 &     88 &   2.92 \\
 SPRC\,56&  13.2 &   11.6 &   120 &   16.0 &    5.6 &   338 &  0.055 &     86 &     48 &   1.21 \\
 SPRC\,57&   3.6 &    2.6 &   145 &    5.2 &    1.8 &    27 &  0.070 &     86 &     56 &   1.45 \\
 SPRC\,58&   7.6 &    7.2 &   103 &   24.0 &    3.2 &   332 &     -- &     85 &     70 &   3.16 \\
 SPRC\,59&   4.5 &    3.5 &    70 &   12.3 &    2.6 &   343 &     -- &     78 &     82 &   2.73 \\
 SPRC\,60&   3.4 &    3.2 &   135 &    9.9 &    2.4 &    55 &  0.078 &     73 &     80 &   2.94 \\
\hline
\end{tabular}
\end{table*}

\setcounter{table}{0}

\begin{table*}
\caption{(continue)}
\begin{tabular}{l|r|r|r|r|r|r|r|r|r|r}\hline\hline
Name    & \multicolumn{3}{c|}{Inner disk}	& \multicolumn{3}{c|}{ Polar ring} &     & &    &   \\	
   	    &a, $''$&b, $''$ &PA, $\degr$&a, $''$&b, $''$ &PA, $\degr$& z & $\delta_1$, $\degr$& $\delta_2$, $\degr$ & $D_{ring}/D_{disk}$\\	
\hline
 SPRC\,61&   8.4 &    5.2 &    49 &   12.0 &    2.4 &   320 &  0.046 &     82 &     83 &   1.43 \\
 SPRC\,62&   2.6 &    2.4 &    45 &    5.2 &    1.7 &    45 &     -- &     48 &     86 &   2.00 \\
 SPRC\,63&   2.4 &    1.8 &   130 &    4.3 &    1.4 &    60 &  0.074 &     62 &     88 &   1.79 \\
 SPRC\,64&   3.1 &    2.2 &   350 &    5.2 &    1.6 &   101 &     -- &     87 &     62 &   1.67 \\
 SPRC\,65&  11.9 &    2.6 &    92 &    9.9 &    3.2 &   310 &  0.067 &     44 &     39 &   0.83 \\
 SPRC\,66&   8.8 &    4.4 &    33 &    9.2 &    2.8 &   117 &  0.087 &     76 &     86 &   1.05 \\
 SPRC\,67&  18.6 &    4.4 &   113 &   25.7 &    3.2 &   347 &  0.028 &     55 &     53 &   1.38 \\
 SPRC\,68&  15.8 &    4.8 &    30 &    7.9 &    4.8 &   115 &     -- &     78 &     86 &   0.50 \\
 SPRC\,69&   5.9 &    3.6 &   135 &   14.7 &    2.8 &    32 &  0.025 &     85 &     73 &   2.47 \\
 SPRC\,70&   5.2 &    2.4 &   100 &    9.1 &    3.0 &    13 &  0.069 &     79 &     84 &   1.77 \\
SPRC\,241&  13.9 &    6.7 &     9 &   27.7 &   15.8 &   110 &  0.018 &     83 &     66 &   2.00 \\
SPRC\,260&  16.6 &    7.9 &    82 &   27.7 &   15.8 &    15 &  0.021 &     57 &     87 &   1.67 \\
  PRC\,A3&  60.0 &   32.0 &    38 &   46.0 &   18.0 &   110 &  0.003 &     63 &     87 &   0.77 \\
  PRC\,A4&  14.3 &    7.9 &   133 &   43.6 &    4.4 &    53 &  0.023 &     78 &     84 &   3.05 \\
  PRC\,A6&  15.8 &    7.9 &   125 &   49.5 &    7.9 &    17 &  0.018 &     78 &     69 &   3.13 \\
  PRC\,B9&  15.8 &   10.7 &     5 &   21.4 &    5.9 &    69 &  0.020 &     60 &     82 &   1.35 \\
 PRC\,B17&  26.0 &   10.0 &   135 &   32.0 &    6.0 &    24 &  0.004 &     74 &     66 &   1.23 \\
 PRC\,C13& 122.8 &   39.2 &    40 &  277.2 &   59.4 &   352 &  0.003 &     46 &     54 &   2.26 \\
\hline
\end{tabular}
\end{table*}

Knowing only $\rm PA$ and $i$ for the two planes, with the angle
$i$ measured by formula (\ref{eq1}), we cannot
unambiguously determine the angle therebetween. Additional
information as of which side of the disk and the ring is
nearest to the observer is yet required. This can be
conceived from the distribution of the dust lanes in the image of
the galaxy, but in the vast majority of studied objects their
angular size is too small compared to the resolution of SDSS
images. In general, there are two solutions for the possible angle
between the components of the ring and the
galaxy~\citep{Moiseev2008}. These values are denoted
below as \linebreak $\delta_1$ and $\delta_2$:
\begin{equation}
\cos \delta_{1,2} =  \pm\cos ({\rm PA}_0-{\rm PA}_1) \sin i_0 \sin i_1  
 +\cos i_0 \cos i_1.
\label{eq2}
\end{equation}
Here ${\rm PA}_0$ is the position angle of the central galaxy,
${\rm PA}_1$ is the position angle of the polar ring,  $i_0$ and
$i_1$  are the angles of inclination of the galaxy and of the
polar ring to the line of sight, respectively. The results are
listed in the table~1.

Figure~\ref{fig_2} shows the distributions of the angle
between the outer ring and the disk for our sample. The left
histogram shows the distribution for all possible values of
$\delta$ from~(\ref{eq2}). Despite the uncertainty in
the estimates of $\delta$, the outer ring structures mostly prove
to be oriented close to the polar plane:  $\delta\ge70\degr$. This
means that the polar configuration in PRGs occurs much more
frequently than the inclined one. This is consistent with the
theoretical studies, which reveal the existence of stable orbits
in the case of axially symmetric (but not spherical) or triaxial
distribution of gravitational potential in the polar plane. At the
same time, the orbits in the plane notably different from polar
prove to be unstable: the stellar ring is destroyed by the
differential precession~\citep{Sparke1986}, and the gas
ring rather quickly (in several revolutions) settles into one of
the principal planes of the central galaxy  \citep[see, e.g.][]{HabeIkeuchi1985}. The lifetime
and the   evolution of the inclined ring depends on
the shape of distribution of the potential
\citep[][and references therein]{Peletier1993,Ideta2000}.

If we, in agreement with the above theoretical studies,
assume that out of the two possible values  $\delta_1$ and
$\delta_2$ the most likely is the angle closer to $90\degr$, then
the distribution of inclination angles of rings to the disk will
take the shape shown in Fig.~\ref{fig_2} to the right.
The peak of the distribution becomes considerably narrower, in
such a way that for 95\% of all sample objects $\delta>70\degr$.
Two important points should be noted.
\begin{itemize}
\item Even with this ``optimistic'' assumption about the
implementation of cases of large angles only,  6\% of the sample
(five objects) are still galaxies with moderately inclined
external structures:  \mbox{$\delta=40$--$55\degr$}.  It is
possible that they are not stable and we have caught them in the
process of destruction (sedimentation in a disk plane). Detailed
modeling of the characteristics of these specific galaxies may
yield the answer. 

\item A precisely polar orientation (within
$5\degr$ from the plane orthogonal to the disk) is observed only
in 39 galaxies.
\end{itemize}

It should of course be borne in mind that the calculation of
angles of inclination via formula~(\ref{eq1}) implies
that the polar structures are flat and round. At the same time,
there are several indications that, firstly, polar rings must
possess weak internal ellipticity~\citep{Iodice2003}
and, secondly, the structures deviating from the polar plane have
to warp~\citep{Sparke1986}, and hence it is not always
correct to attribute them a single angle. However, we believe that
the above-mentioned statistical regularities should be explained
within the theoretical calculations.

In general, the given distributions by the angle $\delta$ are
similar to Fig.~3 from the earlier paper by
\citet{Whitmore1991}, where the author also
makes a conclusion that the observed distribution of the
inclination angle of the ring to the disk is explained by  the
preference of the polar spatial orientation. At the same time,
analyzing a few times larger sample of PRGs, we cannot agree with
the conclusion of~\citet{Whitmore1991} that the absence
of rings with a moderate inclination (in our terms it means
$\delta<55\degr$) is only due to the selection effect. On the one
hand, indeed, during the visual search of PRG candidates, the
objects in which both components are visible at a large angle to
the line of sight prove to be more noticeable: the average value
of the angle $i_0$ in our sample is $56\degr$. Moreover, polar
structures are visible at a yet larger angle: the average $i_1$ is
$70\degr$, given that $i_1<60\degr$ only for eight rings. However,
even if we only select the objects visible edge-on, then
from~(\ref{eq2}) for $i_0\approx i_1\approx90\degr$ we
get \mbox{$\cos\delta\approx\cos({\rm PA}_0-{\rm PA}_1)$}. Hence,
moderately inclined rings would attract attention by the
corresponding difference in the position angles. Therefore, the
selection effect cannot make the observed distributions of
$\delta$ flatter than those shown in Fig.~\ref{fig_2}.

\subsection{Spatial Scales}

The calculations of linear diameters were performed for \mbox{$H_0
= 71$ km/s/Mpc} from the known redshifts of galaxies, listed in
\citet{Moiseev_SDSS}, where they, in turn, were
taken from the SDSS and NED. Figure~\ref{fig_3} shows a
histogram of distribution of the extrnal diameters of rings.
Although the scatter in their values is quite large, almost all
the rings are smaller than $40$~kpc at an average diameter of
about $20$~kpc. Importantly, since the measurements were done from
the optical images, this is the case of stellar polar structures
only. In neutral gas the size of the outer structures can be much
larger, and the inclination to the plane of the disk may vary. For
example, the distribution of H\,I in the PRC\,A3 (NGC\,2685)
galaxy agrees with the stellar polar ring in the inner regions,
but at large radii it warps the way that it again fits in the
plane of the main galaxy~\citep{Jozsa2009}.

SPRC\,50 largely excels from the total distribution, its image is
shown in Fig.~\ref{fig_1}. The central galaxy is
surrounded by a giant disk of low surface brightness. The spectrum
of the core of SDSS~DR7 reveals absorption features characteristic
of early-type galaxies at redshift \mbox{$z=0.078$} which
corresponds to the scale of $1.5$~kpc/arcsec. If the polar disk is
not a result of a random projection, its diameter reaches
$90$~kpc. This is comparable with the sizes of the most extended
galactic disks known to date, such as \linebreak Malin\,1
\citep[$90$~kpc in the diameter, according][]{Bothun1987}, the prototype of the low surface
brightness galaxies,  or a ring galaxy UGC\,7069 \citep[$115$~kpc,
according to][]{ugc7069}. The existence of such
giant polar structures itself is a serious challenge to the
theories of galaxy formation, unless, of course, spectral
observations confirm that SPRC\,50 is not a random projection but
a real PRG.

\begin{figure}
\centerline{\includegraphics[width=8.5cm]{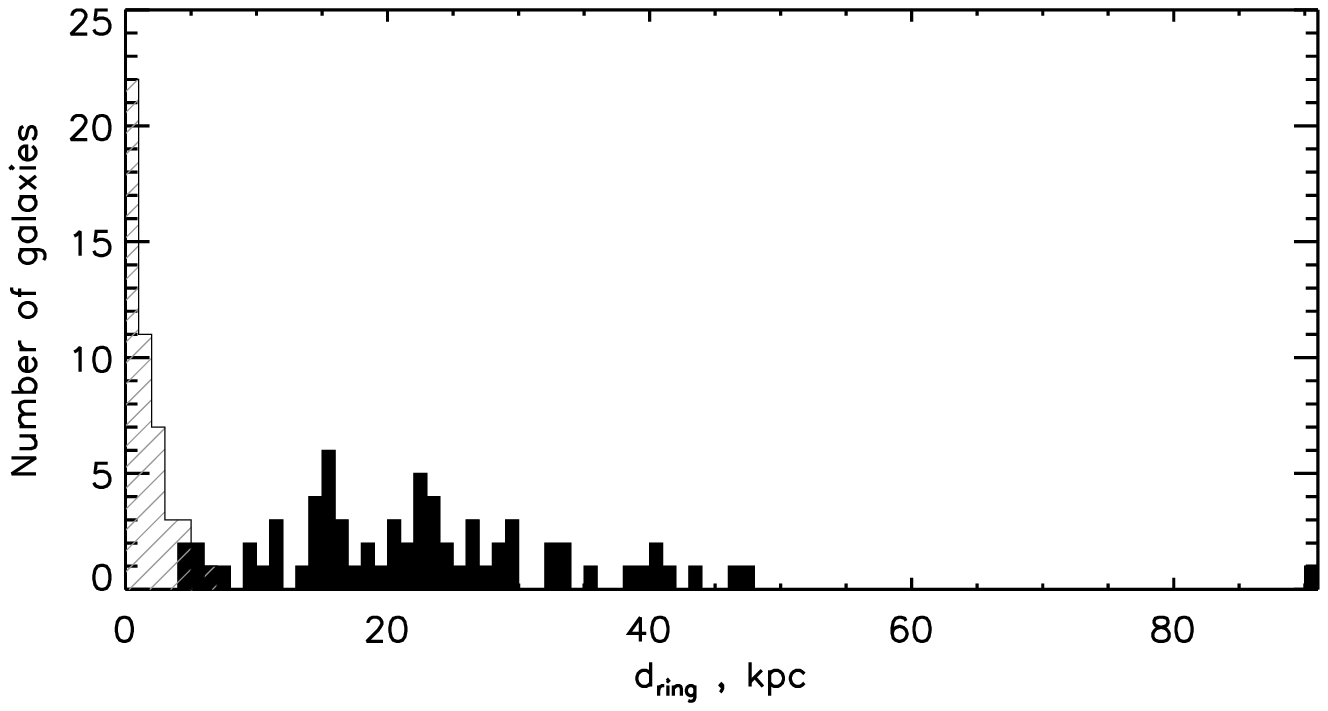}}
\centerline{\includegraphics[width=8.5cm]{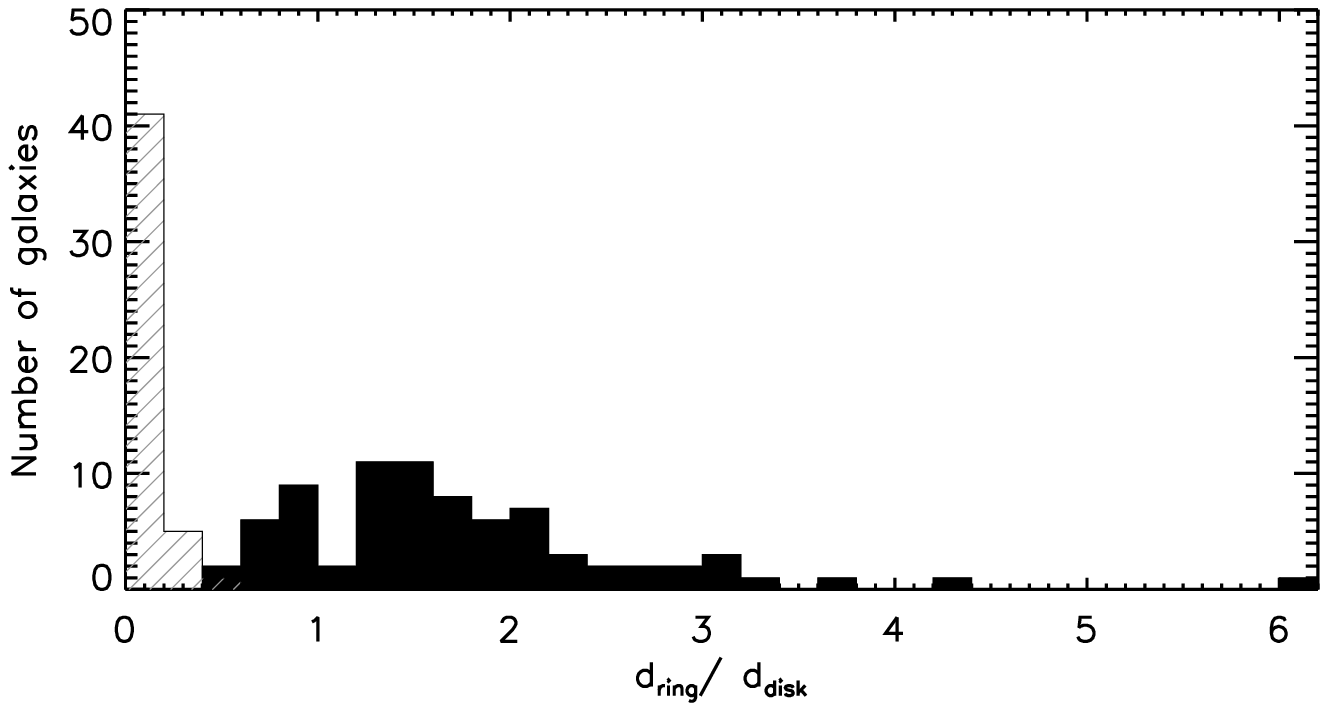}}
\caption{The distributions of polar
structures by the diameter. Top---linear sizes,
bottom---normalization to the diameter of the central disk. Black
color in the figure denotes the outer structures, hatching---the
inner polar structures from~\citet{Moiseev2012}.}.
\label{fig_5}
\end{figure}

Figure~\ref{fig_3} (right) shows the distribution of
the outer diameters of the rings, normalized to the optical
diameter of the galaxy:  $d_{\rm ring}/d_{\rm disk}$. Note that
the outer boundaries of the low brightness regions measured by our
method correspond to the traditional criterion of the size of the
disk by the surface brightness level of $25^{\rm m}$ per a square arcsec ($D_{25}$).
For the majority of galaxies, the diameter of the ring does not
exceed from three to four diameters of the disk, with the average
of about \mbox{$d_{\rm ring}/d_{\rm disk}\approx1.7$.} In this
distribution there is a galaxy with an unusually large relative
size of the ring, SPRC\,42. Its image is shown in
Fig.~\ref{fig_1} to the right. The diameter of the
polar component, visible edge-on with slightly warped outer parts
makes up about $47$~kpc. Such a long structure surrounding a
relatively small lenticular galaxy should be more correctly called
the ``polar disk.'' We are now conducting additional
observations of this intriguing object in the optical and radio
bands.

\begin{figure*}
\centerline{\includegraphics[width=12.5cm]{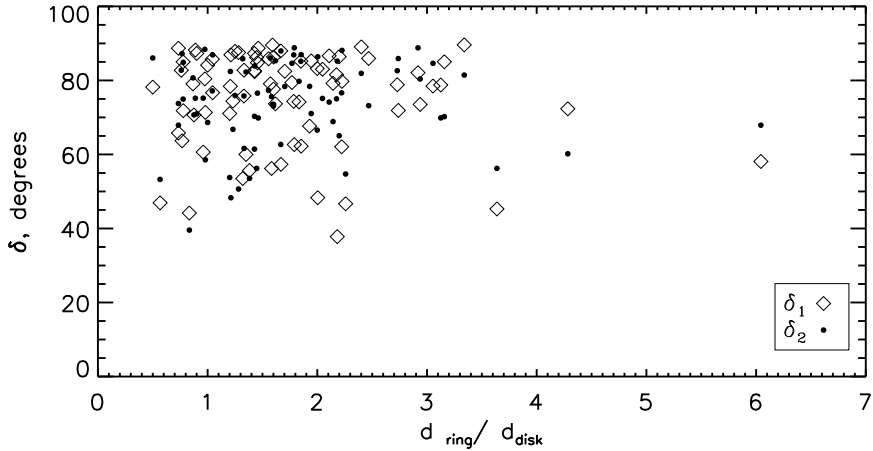}}
\caption{Dependence of the inclination
angles on the diameter of outer polar structures, normalized to
the size of the galaxy. The angles $\delta_1$, $\delta_2$ are
marked by different symbols.} \label{fig_new}
\end{figure*}

In 22\% of the sample (17 galaxies)   $d_{\rm ring}/d_{\rm
disk}\le1$. The entire bright part of these objects' polar rings
rotates crossing the disk of the host galaxy. It is clear that the
spatial density of stars is small and direct collisions of stellar
components do not occur. But if the polar ring contains gas, shock
waves may be generated while the gas clouds pass through the
gravitational well of the stellar disk. Moreover, if the inner
disk contains its own gas, direct collision of gas flows becomes
possible. Unfortunately, literature has almost no theoretical
considerations of the process, with the exception of one article
by \citet{Wakamatsu1993}. It is possible that
exactly such a picture, i.e., impact ionization of gas on inclined
orbits is observed in NGC\,7743. This is a lenticular galaxy, in
which the entire ionized gas at \mbox{$r=1.5$--$5.4$ kpc} is
located in the plane inclined by $34\degr$ or $77\degr$ to the
stellar disk~\citep{Katkov2011_N7743}.

It is interesting to compare the distributions in Fig.~3 with
those for the inner polar rings and disks observed in the
circumnuclear regions of a number of nearby galaxies.
Figure~\ref{fig_5}, in addition  to the already
considered distributions of the diameters and relative diameters
of  the outer polar structures, shows the data on 47 galaxies with
confirmed inner structures from~\citet{Moiseev2012}. For
the inner polar structures, the diameter of the disk of the galaxy
was accepted as the diameter of $D_{25}$  taken from the NED
database. We can see that the sizes of inner and outer polar
structures form a continuous sequence from a hundred parsecs to
tens of kiloparsecs. By the values of normalized diameters both
samples come into contact at $d_{\rm ring}/d_{\rm
disk}\approx0.5$, hence the galaxy NGC\,5014 (SPRC\,40) is
included in both lists: inner structures and SPRC catalog. At the
same time, a relative shortage of  ``intermediate size'' polar
structures  ($d_{\rm ring}/d_{\rm disk}\approx0.4$--$0.7$) is
noticeable. Only two galaxies from the SDSS sample find themselves
here: SPRC\,40 and SPRC\,68.

Such a bimodal distribution of the relative size of polar
structures is most likely due to the fact that in all the cases
considered, the stability of the polar orbit is controlled by the
spheroidal or triaxial gravitational potential. But if the
gravitational potential of the dark halo is important for the
stability of large-scale structures, the bulge (including the
triaxial one) or the central bar play the dominant role for the
inner rings and disks. In the intermediate zone of radial scales,
the shape of the potential  is markedly different from the
spheroidal/triaxial one, since a significant contribution in the
gravitational potential is brought here by the flat stellar disk.
Therefore, the polar/inclined orbits cease to be stable.
Collisions occurring while the disk is crossed by the matter on
polar orbits, mentioned in the previous chapter, is the
manifestation of the same effect. Apparently it is not a
coincidence that in the SPRC\,40 the ``polar'' structure of
intermediate size is strongly inclined to the galactic disk and is
possibly not stationary.

Figure~\ref{fig_new} shows the distribution of points
in the plane   \mbox{($d_{\rm ring}/d_{\rm disk}$, $\delta$).}
Although the scatter of the observed values is quite large, it is
indeed clear that the deviation from the polar plane of
($\delta=90\degr$) is more frequently observed for the rings
crossing the central disk: \mbox{$d_{\rm ring}/d_{\rm disk}<1$.}
The largest rings are as well revealing the drifts from the polar
plane \mbox{($d_{\rm ring}/d_{\rm disk}>3.5$)} which is apparently
related to the development of warp instability at decreasing halo
density.

\section{CONCLUSIONS }
\label{sec_conclusion}

We have considered a sample of 78 most reliable polar ring galaxy
candidates, represented in the SDSS survey, some of which already
have kinematic confirmations. Most of the outer structures are
indeed polar, i.e., inclined at an angle of more than $70\degr$ to
the central plane of the galaxy. We show that the distribution of
relative sizes of the outer polar rings and circumnuclear polar
disks (inner rings) is bimodal, with a ``drop'' falling at the
interval of \mbox{$0.4< d_{\rm ring}/d_{\rm disk} < 0.7$}. This is
most likely due to the fact that polar and inclined structures of
intermediate sizes prove to be short-lived. At that, the stability
of inner polar rings and disks is supported by the gravity of the
bulge, while the outer structures are maintained by the gravity of
the dark halo.

\begin{acknowledgements}
This work was supported by the grants of\linebreak the Russian
Foundation for Basic Research\linebreak (project
\mbox{no.~13-02-00416-a}) and the ``Active Processes in Galactic
and Extragalactic Objects'' basic research program of the
Department Physical Sciences of the RAS OFN-17.  A.~Moiseev thanks
the non-profit Dynasty foundation for the support. In the process
of work we used the NASA/IPAC (NED) database of extragalactic
data, managed by the Jet Propulsion Laboratory of the California
Institute of Technology under the contract with the NASA (USA). 

Funding for the SDSS has been provided by the Alfred P. Sloan
Foundation, the Participating Institutions, the National Science
Foundation, the U.S. Department of Energy, the National
Aeronautics and Space Administration, the Japanese Monbukagakusho,
the Max Planck Society, and the Higher Education Funding Council
for England. The SDSS Web Site is http://www.sdss.org/. 

The SDSS is managed by the Astrophysical Research Consortium (ARC)
for the Participating Institutions. The Participating Institutions
are The University of Chicago, Fermilab, the Institute for
Advanced Study, the Japan Participation Group, The Johns Hopkins
University, Los Alamos National Laboratory, the
Max-Planck-Institute for Astronomy (MPIA), the
Max-Planck-Institute for Astrophysics (MPA), New Mexico State
University, University of Pittsburgh, Princeton University, the
United States Naval Observatory, and the University of Washington. 

\end{acknowledgements}

\end{document}